\documentclass[journal]{IEEEtran}
\ifCLASSINFOpdf
  \usepackage[pdftex]{graphicx}
\else
  \usepackage[dvips]{graphicx}
\fi
\usepackage{amsmath}
\usepackage{array}
\ifCLASSOPTIONcompsoc
  \usepackage[caption=false,font=normalsize,labelfont=sf,textfont=sf]{subfig}
\else
  \usepackage[caption=false,font=footnotesize]{subfig}
\fi
\begin{document}
%
% paper title
% Titles are generally capitalized except for words such as a, an, and, as,
% at, but, by, for, in, nor, of, on, or, the, to and up, which are usually
% not capitalized unless they are the first or last word of the title.
% Linebreaks \\ can be used within to get better formatting as desired.
% Do not put math or special symbols in the title.
% \title{Natural image identification with brain inner-state combined decoding framework}
\title{An encoding framework with brain inner state for natural image identification}
%
%
% author names and IEEE memberships
% note positions of commas and nonbreaking spaces ( ~ ) LaTeX will not break
% a structure at a ~ so this keeps an author's name from being broken across
% two lines.
% use \thanks{} to gain access to the first footnote area
% a separate \thanks must be used for each paragraph as LaTeX2e's \thanks
% was not built to handle multiple paragraphs
%

\author{Hao~Wu, Ziyu~Zhu, Jiayi~Wang,
	    Nanning~Zheng,~\IEEEmembership{Fellow,~IEEE,}
        and Badong~Chen*,~\IEEEmembership{Senior~Member,~IEEE}
\thanks{H. Wu, Z. Zhu, J. Wang, N. Zheng and B. Chen (corresponding author) are with the Institute of Artificial Intelligence and Robotics, Xi'an Jiaotong
	University, Xi'an, Shaanxi 710049, P.R. China. E-mail: xuan.zhi, zhuzy\_2016, jiayiwang@stu.xjtu.edu.cn, nnzheng, chenbd@mail.xjtu.edu.cn}% <-this % stops a space
}

% note the % following the last \IEEEmembership and also \thanks - 
% these prevent an unwanted space from occurring between the last author name
% and the end of the author line. i.e., if you had this:
% 
% \author{....lastname \thanks{...} \thanks{...} }
%                     ^------------^------------^----Do not want these spaces!
%
% a space would be appended to the last name and could cause every name on that
% line to be shifted left slightly. This is one of those "LaTeX things". For
% instance, "\textbf{A} \textbf{B}" will typeset as "A B" not "AB". To get
% "AB" then you have to do: "\textbf{A}\textbf{B}"
% \thanks is no different in this regard, so shield the last } of each \thanks
% that ends a line with a % and do not let a space in before the next \thanks.
% Spaces after \IEEEmembership other than the last one are OK (and needed) as
% you are supposed to have spaces between the names. For what it is worth,
% this is a minor point as most people would not even notice if the said evil
% space somehow managed to creep in.

% The paper headers
\markboth{IEEE TRANSACTIONS ON COGNITIVE AND DEVELOPMENTAL SYSTEMS}%
{An encoding framework with brain inner state for natural image identification}
% The only time the second header will appear is for the odd numbered pages
% after the title page when using the twoside option.
% 
% *** Note that you probably will NOT want to include the author's ***
% *** name in the headers of peer review papers.                   ***
% You can use \ifCLASSOPTIONpeerreview for conditional compilation here if
% you desire.

% If you want to put a publisher's ID mark on the page you can do it like
% this:
%\IEEEpubid{0000--0000/00\$00.00~\copyright~2015 IEEE}
% Remember, if you use this you must call \IEEEpubidadjcol in the second
% column for its text to clear the IEEEpubid mark.

% use for special paper notices
%\IEEEspecialpapernotice{(Invited Paper)}

% make the title area
\maketitle

% As a general rule, do not put math, special symbols or citations
% in the abstract or keywords.
\begin{abstract}
Neural encoding and decoding, which aim to characterize the relationship between stimuli and brain activities, have emerged as an important area in cognitive neuroscience. Traditional encoding models, which focus on feature extraction and mapping, consider the brain as an input-output mapper without inner states. In this work, inspired by the fact that human brain acts like a state machine, we proposed a novel encoding framework that combines information from both the external world and the inner state to predict brain activity. The framework comprises two parts: forward encoding model that deals with visual stimuli and inner state model that captures influence from intrinsic connections in the brain. The forward model can be any traditional encoding model, making the framework flexible. The inner state model is a linear model to utilize information in the prediction residuals of the forward model. The proposed encoding framework can achieve much better performance on natural image identification from fMRI response than forward-only models. The identification accuracy will decrease slightly with the dataset size increasing, but remain relatively stable with different identification methods. The results confirm that the new encoding framework is effective and robust when used for brain decoding.
\end{abstract}

% Note that keywords are not normally used for peerreview papers.
\begin{IEEEkeywords}
connectivity, decoding, fMRI, perception, voxel-wise encoding.
\end{IEEEkeywords}

% For peer review papers, you can put extra information on the cover
% page as needed:
% \ifCLASSOPTIONpeerreview
% \begin{center} \bfseries EDICS Category: 3-BBND \end{center}
% \fi
%
% For peerreview papers, this IEEEtran command inserts a page break and
% creates the second title. It will be ignored for other modes.
\IEEEpeerreviewmaketitle

\section{Introduction}
% The very first letter is a 2 line initial drop letter followed
% by the rest of the first word in caps.
% 
% form to use if the first word consists of a single letter:
% \IEEEPARstart{A}{demo} file is ....
% 
% form to use if you need the single drop letter followed by
% normal text (unknown if ever used by the IEEE):
% \IEEEPARstart{A}{}demo file is ....
% 
% Some journals put the first two words in caps:
% \IEEEPARstart{T}{his demo} file is ....
% 
% Here we have the typical use of a "T" for an initial drop letter
% and "HIS" in caps to complete the first word.
\IEEEPARstart{O}{ne} of the most important goals of cognitive neuroscience is to understand how mental operations are performed in the brain. Characterizing the relationship between stimulus features and brain activities is an effective approach to the goal. This relationship can be explored from two distinct
but complementary perspectives: encoding and decoding. Encoding uses the information of stimuli or tasks to predict brain activities while decoding takes brain activities as input to predict or identify features of stimuli \cite{naselaris_encoding_2011}. Encoding and decoding allow researchers to focus on the storage of mental content in brain regions, rather than on overall levels of activation \cite{haynes_multivariate_2011}.

Functional magnetic resonance imaging (fMRI) provides a powerful tool for measuring human brain activity. Despite the non-invasive property, fMRI offers a rather complicated window on neural activation. First, fMRI measures changes in blood oxygen level dependent (BOLD) caused by neural activity, rather than the activity itself \cite{ogawa_finding_2012}. Thus, it only provides an indirect measure of neural activity. Second, fMRI can cover the entire brain and simultaneously collect responses from hundreds of thousands of individual voxels, with each voxel containing tens of thousands of neurons.

Brain encoding with fMRI usually aims at modeling responses of each individual voxel using sensory, cognitive or task information. Specifically, typical encoding models consist of several distinct components \cite{naselaris_encoding_2011}. The first is a group of stimuli adopted in the experiment. The stimuli set can not only be sensory materials such as visual scenes, movies, or audio stories \cite{kendrick_n._kay_identifying_2008,vu_encoding_2011,nishimoto_reconstructing_2011,naselaris_voxel-wise_2015}, but also can be mental materials such as visual imagery \cite{huth_natural_2016}. The second component is a set of features that characterize the abstract relationship between stimuli and responses. For example, phase-invariant Gabor wavelets were chosen as low level features of image in many visual encoding studies \cite{kendrick_n._kay_identifying_2008}. Natural scene categories were adopted as semantic level features \cite{naselaris_bayesian_2009}. In addition to these hand-designed features, recent developments in machine learning also promoted some unsupervised automatic feature extraction methods, such as sparse coding and deep neural networks \cite{guclu_unsupervised_2014,gucclu2015deep}. The third component is a set of voxel, always selected from one or more regions of interest (ROI) in the brain. The final component is a model that is used to map the features to the responses of each selected voxel. On the other hand, brain decoding is to determine how much can be learned about the world by observing BOLD activity. In particular, decoding model can be used to classify the specific class of the stimulus given the correspondent voxel activity pattern (classification), identify the correct stimulus from a set of novel stimuli (identification), or even reconstruct the details of the stimulus (reconstruction) \cite{kendrick_n._kay_identifying_2008}. Most of the decoding studies used multi-voxel pattern analysis (MVPA) such as SVM classifier, Bayesian learning method, or neural networks to classify or reconstruct the perceived visual stimuli \cite{kamitani_decoding_2005,wen_neural_nodate,zafar_decoding_2017,yu_novel_2018,du_reconstructing_2018}. 

In addition to MVPA approaches, the voxel-wise encoding model can also be converted to decoding models that identify stimulus features from the BOLD activities evoked by the stimulus. This procedure typically consists of four stages. First, train an encoding model for each voxel on the training data. Then choose a group of voxel that have high predictive power. Third, build a large image set, and predict voxel responses for each image. Finally, calculate similarities between the predicted response patterns and a measured response pattern in the testing data, and choose the image with the highest similarity as the identification result \cite{kendrick_n._kay_identifying_2008}, or choose images with relatively higher similarity to reconstruct the true image correspond to the given response pattern \cite{naselaris_bayesian_2009,nishimoto_reconstructing_2011}. Comparing with MVPA, the voxel-wise encoding model offers some important advantages. It resolves geometric and representational ambiguity which is inherent to MVPA \cite{naselaris_resolving_2015}.

\begin{figure}[!t]
\centering
\includegraphics[width=3 in]{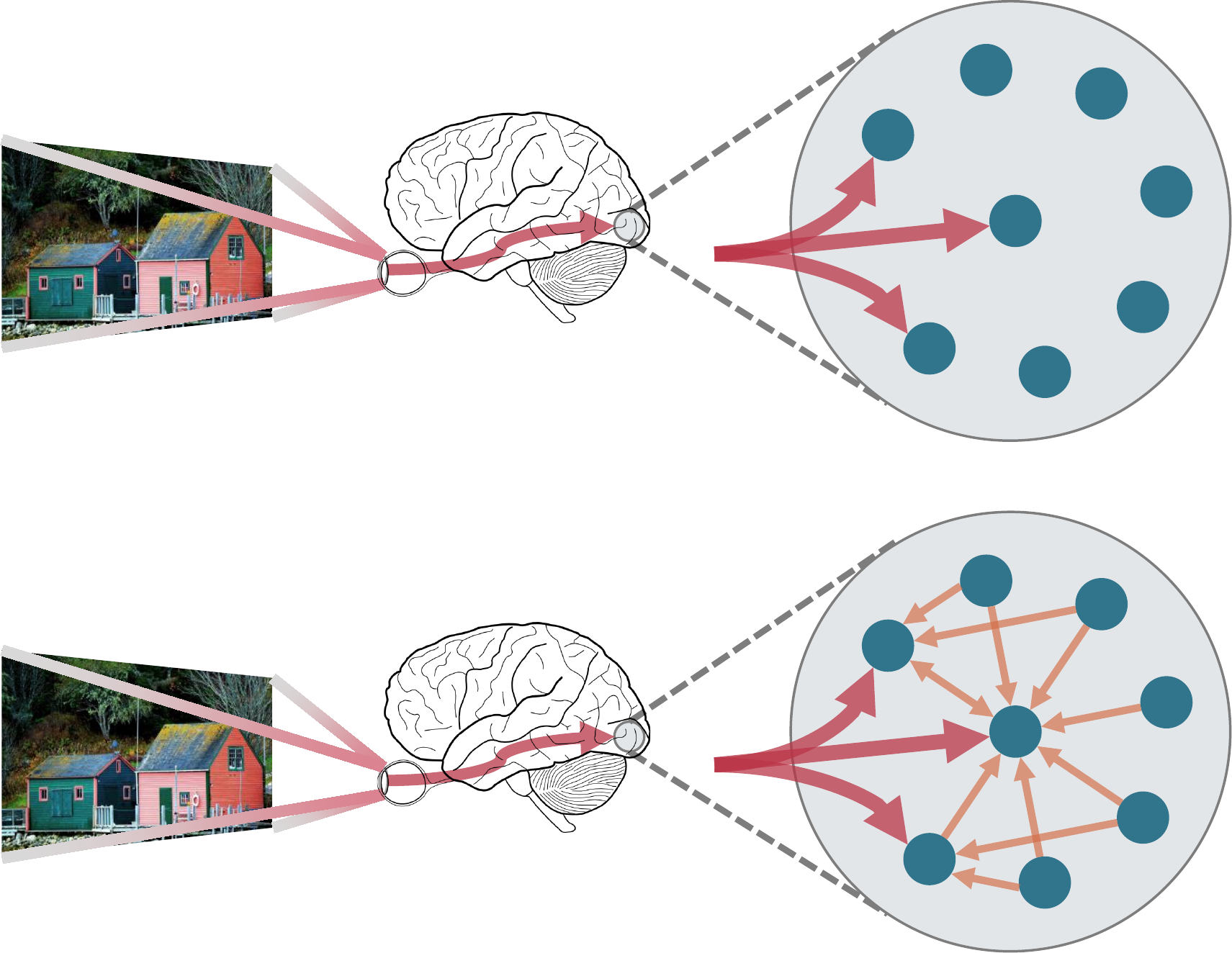}
\caption{Schematic of visual encoding. Cyan dots in gray circles represent voxels in visual areas. Red arrows indicate flows of stimulus information, while orange arrows indicate flows of non-stimulus information between voxels. Conventional voxel-wise encoding model assumes that brain activity fluctuates only with external stimuli (top panel), which ignores abundant connections between voxels (bottom panel).}
\label{fig_encoding_compare}
\end{figure}

Despite the many advantages of the voxel-wise encoding model, there are still some limitations to it. The most crucial one is that the explained variance ($R^2$) of voxel activity remains low, regardless whether the extracted features and encoding mappings are simple or complex. $R^2$ is defined as the proportion of the variance in the dependent variable (voxel response) that is predictable from the independent variables (stimulus features), which is usually used as a measurement of encoding accuracy of voxel-wise encoding model. We summarized some recent fMRI encoding studies that covered various encoding models or feature extraction approaches, from simple linear model to complicated non-linear model, from hand-designed to unsupervised learned features, and listed their reported encoding $R^2$ (either mean or maximum) in Table \ref{table_R2list}. The maximum $R^2$ is 0.65 with the mean $R^2$ is 0.28, which means that all these encoding models only captured a little part of the fluctuation in voxel activities. The unexplained part in brain activities is considered as noises. Another limitation of voxel-wise encoding model is that it ignores the fact that voxels are interconnected. The voxel-wise encoding model finds a special mapping from stimulus features to BOLD responses for each individual voxel, thus it can be called as forward model. The logic behind a forward model is that human brain acts like a camera (for visual sense), whose activity changes are merely a reflection of fluctuations in the external world (Fig \ref{fig_encoding_compare}, top panel). However, the brain functions much more beyond a simple sensor, it's more appropriate to see it as a state machine, which not only receives input signals from sensory organs, but also maintains a very large amount of internal states at every moment \cite{vanderVelde1993,vidaurre_discovering_2017}. Many studies have demonstrated that the early visual cortex receives information from lateral geniculate nucleus (LGN) (forward connection), as well as from intra-area (lateral connection) and high level areas (feedback connection) (Fig \ref{fig_encoding_compare}, bottom panel) \cite{gilbert_top-down_2013,williams_feedback_2008,shibata_perceptual_2011}. All these facts show that it is not enough to consider only the forward signal in the current voxel-wise encoding model.

\renewcommand{\arraystretch}{1.5}
\begin{table*}[!t]
%% increase table row spacing, adjust to taste
%\renewcommand{\arraystretch}{1.3}
% if using array.sty, it might be a good idea to tweak the value of
% \extrarowheight as needed to properly center the text within the cells
\caption{$R^2$ of different encoding models}
\label{table_R2list}
\centering
%% Some packages, such as MDW tools, offer better commands for making tables
%% than the plain LaTeX2e tabular which is used here.
\begin{tabular}{c|c|c|c|c|c}
\hline
Study & Statistic method & $R^2$ & Encoding model & Brain area & Stimulus type\\
\hline
Guclu et al.\cite{guclu_unsupervised_2014} & mean & 0.28 & unsupervised feature learning & V1 & natural image\\
%\hline
Guclu et al.\cite{gucclu2015deep} & mean & 0.25 & deep neural networks & V1 & natural image\\
%\hline
Vu et al.\cite{vu_encoding_2011} & maximum & 0.65 & non-linear model & V1 & natural image\\
%\hline
Naselaris et al.\cite{naselaris_bayesian_2009} & mean & 0.30 & semantic encoding model & anterior to lateral occipital & natural image\\
%\hline
Huth et al.\cite{huth_natural_2016} & maximum & 0.36 & embedding semantic feature & whole brain & audio story\\
%\hline
Nishimoto et al.\cite{nishimoto_reconstructing_2011} & mean & 0.16 & motion energy model & early visual area & movie\\
%\hline
Naselaris et al.\cite{naselaris_voxel-wise_2015} & maximum & 0.36 & Gabor wavelet pyramid  & V1 and V2 & mental image\\
\hline
\end{tabular}
\end{table*}

%The reason why voxel-wise encoding model couldn't achieve high prediction accuracy have been discussed a lot. The most important one is that fMRI data are noisy. Noise in fMRI comes from several sources, include physical factors related to MR imaging (such as thermal noise), physiological factors (such as respiration), and various cognitive factors (such as arouse, attention, and memory) \cite{naselaris_encoding_2011,buxton_modeling_2004}. 

The resting state fMRI has provided us with a paradigm that makes use of interconnection information between voxels. In the absence of external stimulus or task, the default mode network of the brain can be investigated by studying the correlation between activities in brain regions \cite{greicius_functional_2003, greicius_resting-state_2009}. However, it remains difficult to use the interconnection information in the scenario of voxel encoding in which case the brain activity is a mixture of internal fluctuations and responses evoked by external stimuli. Take visual perception as an example, the natural scene seen by the eyes has its intrinsic structure, such as spatial correlation (neighboring spatial locations are strongly correlated in intensity), color distribution (the light falling on an image at a given location has a spectral distribution), and high-order statistical features \cite{bruno_a._olshausen_natural_2001,oliva_role_2007}. Given stimuli with highly correlated intrinsic structure, it is hard to determine whether the correlation between voxel activities is caused by the stimuli or by the intrinsic connection between voxels.

Based on the fact that neural activity is not only related to external stimuli, but also to the state of internal connections in the brain, we suggest that even without the different kinds of noises which are conventionally thought as the main reason for bad encoding performance, a forward-only encoding model could not achieve high predictive performance. On decoding perspective, if the predicted activity pattern of an image by a forward-only model is exactly identical to the measured activity pattern, then choosing that image as the identification result may not be optimal, for totally being predicted by the external world means human brain acts like a simple sensor and has no inner state. In this paper, we proposed an encoding framework that takes account of brain inner-state (ISF). The framework consists of two components. One is a forward encoding model, which deals with responses evoked by external stimulus. The other is an inner-state model, which handles responses that cannot be characterized by external stimulus. The information captures by the inner state model is linearly independent to the visual stimuli, hence better reflect the intrinsic connectivity in the brain. It's worth noting that the forward model in this framework can be any kind of traditional voxel-wise encoding model, which makes it more flexible. To estimate the inner state of one voxel, one needs to know other voxels' true activities, this feature makes the proposed encoding framework naturally suitable for brain decoding, where a measured activity pattern must be given at first. We used a visual fMRI experiment dataset and compared the encoding and decoding performances of the proposed framework with that of several other forward-only encoding models. The results showed that our proposed encoding framework can achieve significantly better performances on both encoding and decoding tasks.
% 我们的模型取得了非常优异的性能

\section{Materials and methods}
%In this section, we start with introducing the forward model which consists of a Gabor wavelet pyramid (GWP) model to estimate the receptive field (RF) of each voxel and a linear regression model to map features of stimulus to voxel activity \cite{Lee2013A,Singer1995Visual}. Then we describe the details of the hidden state model, and show how to use the model to perform decoding work, i.e. to identify which visual stimulus is seen when a pattern of voxels activity is given. Finally, the experimental design and fMRI dataset that we use to validate our model are introduced.

In this section, we start with introducing the brain inner state framework, then several conventional forward encoding models are described in detail, followed by the introduction of how to carry on image identification using encoding models.

\subsection{Brain inner state framework}
%It's obvious that the forward model only use features of stimulus to encode brain response. But there's more and more evidence that shows even early visual cortex does not only process forward information, but also receives signals from lateral and feedback connections (details are shown in Discussions). To utilize this neuroscience fact, we introduce a hidden state that represents the non-stimulated activity for each voxel, then combine stimulus features with the hidden state to predict voxel's response, as shown in Figure \ref{two_en}.

\begin{figure*}[!t]
	\centering
	\subfloat[]{
		\includegraphics[width=6 in]{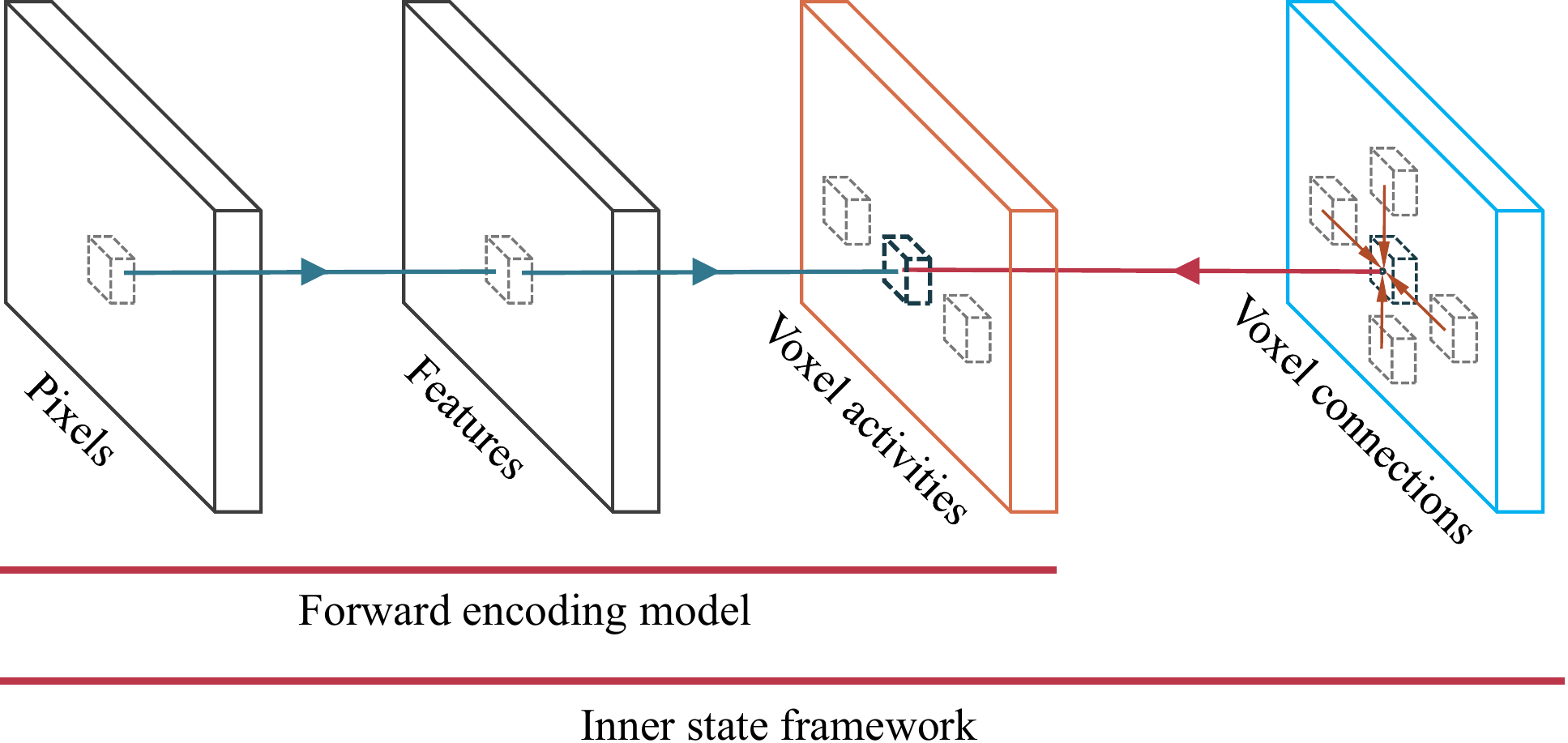}
		\label{fig_framework_encoding}}
	\\
	\subfloat[]{
		\includegraphics[width=6.2 in]{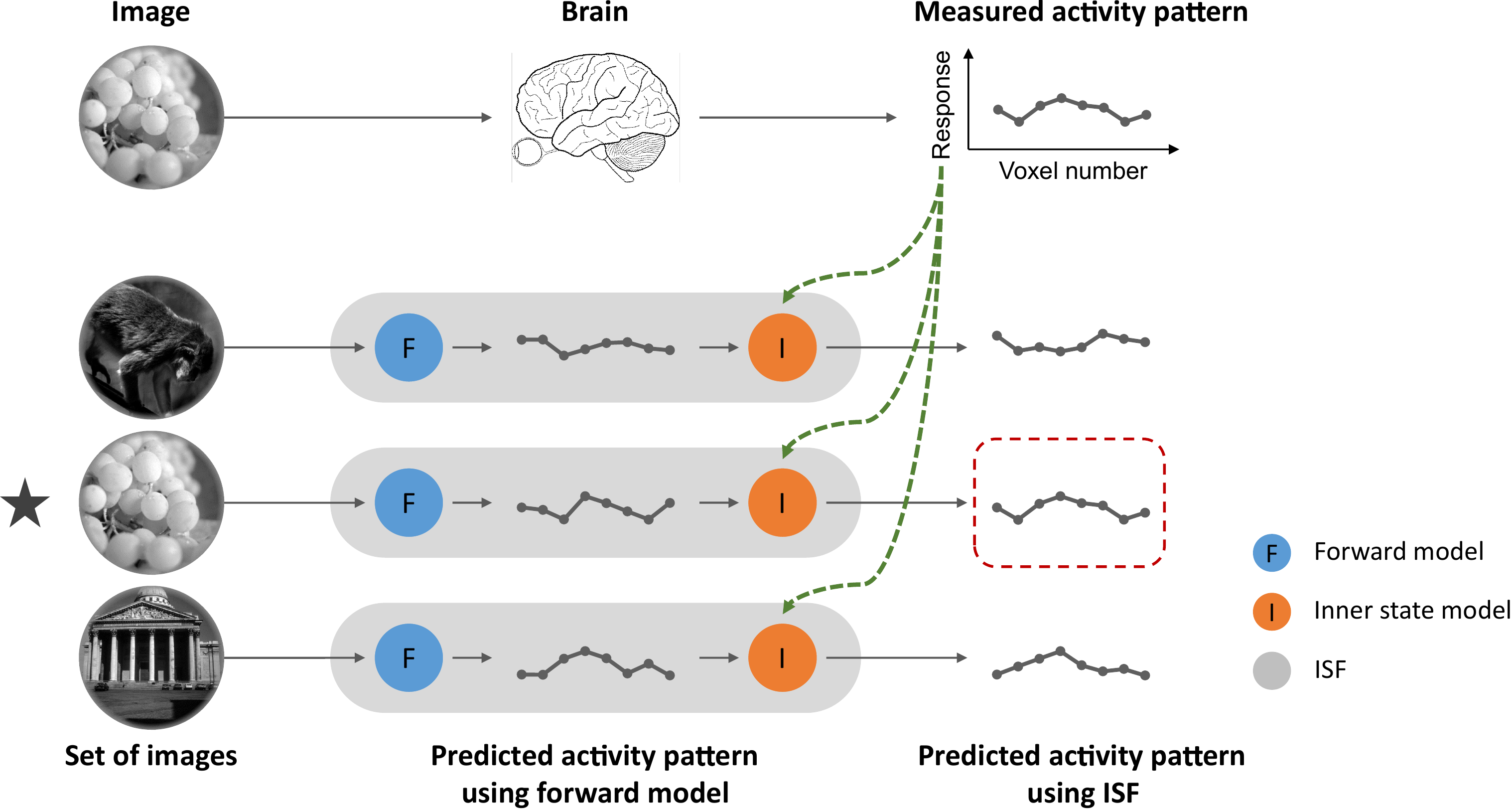}
		\label{fig_framework_decoding}}
	\caption{Schematic diagram of ISF. (a) Encoding framework. (b) Decoding procedure.}
	\label{fig_framework}
\end{figure*}

To predict brain activities under external stimuli, a forward encoding model takes images as input and models the activities as:
\begin{equation}\label{eq_forward_model}
	\boldsymbol{v}_i = \boldsymbol{f_i}(\boldsymbol{X})\boldsymbol{\beta}_i + \boldsymbol{e}_i
\end{equation}
where $\boldsymbol{v}_i$ is the activity vector of the $i$'th voxel, $\boldsymbol{X}$ is the pixel values of stimuli, $\boldsymbol{f_i}$ is some feature extraction functions which convert gray value of pixels into local feature values, $\boldsymbol{\beta}_i$ is a weight vector, and $\boldsymbol{e}_i$ represents the noise term. However, it is not enough to assume that voxel fluctuations are only caused by external stimuli. Besides the determined stimuli and the undetermined noises, there are some structural fluctuations from the intrinsic connections in brain that influence voxels' activity. As shown in Fig. \ref{fig_framework_encoding}, ISF includes an extra inner state model which captures influence from the intrinsic connections compared to the forward-only model. Thus encoding with ISF is defined below:

\begin{equation}\label{eq_isf}
\boldsymbol{v}_i = \boldsymbol{f_i}(\boldsymbol{X})\boldsymbol{\beta}_i + \boldsymbol{s}_i\lambda_i + \boldsymbol{e}_i
\end{equation}
where $\boldsymbol{v}_i$, $\boldsymbol{X}$, $\boldsymbol{f_i}$, $\boldsymbol{\beta}_i$, and $\boldsymbol{e}_i$ are the same as in (\ref{eq_forward_model}). $\boldsymbol{s}_i$ is the inner state associated with the $i$'th voxel, and $\lambda_i$ is the weight.

%\begin{figure*}[!t]
%	\centering
%	\includegraphics[width=6 in]{ISF.eps}
%	\caption{Simulation results for the network.}
%	\label{fig_isf}
%\end{figure*}

How to estimate the inner state for each voxel is one of the main concerns in this work. We proposed a data driven method to accomplish it. First, a forward model is fitted on the training data for each voxel to get a residual matrix with each column one voxel's residual vector:

\begin{equation}\label{eq_residual}
\boldsymbol{\epsilon}_i = \boldsymbol{v}_i - \boldsymbol{f_i}(\boldsymbol{X})\boldsymbol{\tilde{\beta}}_i
\end{equation}

\begin{equation}\label{eq_E}
\boldsymbol{E} = [\boldsymbol{\epsilon}_1,\boldsymbol{\epsilon}_2,\ldots,\boldsymbol{\epsilon}_p]
\end{equation}
where $\boldsymbol{\epsilon}_i$ is the residual vector of the $i$'th voxel using forward model, $\boldsymbol{\tilde{\beta}}_i$ is the estimated weight vector, $\boldsymbol{E}$ is the residual matrix, and $p$ is the total number of voxels. Notice that using the residual matrix instead of the original activity matrix to estimate the inner state is reasonable, since the structural information of natural images contained in the latter can be confusing.

To obtain a connectivity map for each voxel, we calculate the similarity matrix for $\boldsymbol{E}$ using Pearson correlation, then choose a series of voxels whose Pearson's $r$ are above a given threshold as the connected voxels to each voxel, and then estimate the inner state for each voxel using the first principal component of residual vectors of its own connected voxels. The equations are shown below.

\begin{equation}\label{eq_Ei}
\boldsymbol{E}_i = [\boldsymbol{\epsilon}_j,\boldsymbol{\epsilon}_k,\ldots,\boldsymbol{\epsilon}_m]
\end{equation}

\begin{equation}\label{eq_pca}
\begin{aligned}
\tilde{\boldsymbol{s}}_i &= PCA(\boldsymbol{E}_i) \\
&= \boldsymbol{E}_i\boldsymbol{\alpha}_i
\end{aligned}
\end{equation}
where $ \boldsymbol{E}_i $ is residual matrix for the $i$'th voxel, $j$,$k$,$m$ indicate voxels whose residual vectors show high correlation with the $i$'th voxel, and $\boldsymbol{\alpha}_i$ indicates PCA projection vector.

After the inner state is estimated by a linear combination of the residual matrix, the coefficient $\lambda_i$, which represents the impact of the inner state on the activity of voxel $i$, is given by:
\begin{equation}\label{eq_lambda}
\tilde{\lambda}_i = (\tilde{\boldsymbol{s}}_i^T\tilde{\boldsymbol{s}}_i)^{-1}\tilde{\boldsymbol{s}}_i^T\boldsymbol{\epsilon}_i
\end{equation}

$\boldsymbol{\beta}$, $\boldsymbol{\alpha}$ and $\lambda$ are fitted on the training data set for each voxel.

\subsection{Forward encoding models}
To demonstrate our proposed encoding framework can be combined with any kind of effective forward model to improve the encoding and decoding performance, we introduced several forward encoding models, include Gabor wavelet pyramid model, gross local orientation model, and retinotopy-only model. A model that simply predicts a zero response for any stimulus which is called zero model is also introduced here. The predictive power of the four encoding models varies with their complexity. As mentioned in the introduction section, an encoding model consists of two stages: feature extraction and feature mapping. For each model, only its feature extraction method is described in detail, since an encoding model is mainly characterized by it.

\subsubsection{Gabor wavelet pyramid model}
The Gabor wavelet pyramid (GWP) model has long been considered as a standard model of how early visual cortex represents local shape \cite{tai_sing_lee_image_1996,jones_evaluation_1987,daugman_uncertainty_1985}. Previous results suggest that fMRI activity in the primary visual cortex reflects the average activation of a population of Gabor filters \cite{rainer_nonmonotonic_2001}. The GWP model used in the present work describes tuning along the dimensions of orientation, space and spatial frequency.

A 2-D isotropic Gabor wavelet is defined by a sinusoidal wave multiplied by a Gaussian function, as shown below:

\begin{equation}\label{eq_gabor}
G(x,y,\lambda,\theta,\phi,\sigma) = \exp(-\frac{x'^2+y'^2}{2\sigma^2})\cos(2\pi\frac{x'}{\lambda}+\phi)
\end{equation}
where 
\begin{align*}
x' &= x\cos\theta + y\sin\theta \\ 
y' &= -x\sin\theta + y\cos\theta
\end{align*}
and $x, y, \lambda, \theta, \phi, \sigma$ indicate the location, spatial frequency, orientation, phase, and size of the wavelet respectively.

In this work, Gabor wavelets are generated at six spatial frequencies, include 1, 2, 4, 8, 16, and 32 (unit: cycles per field of view). At each spatial frequency $f$ cycles per field-of-view (FOV), wavelets are positioned on an $f \times f$ grid. At each grid position, wavelets occur at eight equally spaced orientations ranged from $0^\circ$ to $157.5^\circ$, and two orthogonal phases: $0^\circ$ and $90^\circ$ (Fig ). Preferred spatial frequency ($1/\lambda$) and size ($\sigma$) are not completely independent: $\sigma = a\lambda$ with $a$ between 0.3 and 0.6 for most cells \cite{daugman_uncertainty_1985}. In the following, we used a typical $a$ value of 0.56 \cite{kendrick_n._kay_identifying_2008}.

Each image is projected onto Gabor wavelets. The response of each orthogonal pair (two orthogonal phases) were squared, summed and square-rooted, reflecting the contrast energy of the wavelet pair. The GWP feature can be expressed as

\begin{equation}
F_{pos,\theta,\lambda} = \sqrt{\sum_{\phi}(stim \cdot G_{pos,\theta,\lambda,\phi})^2}
\end{equation}
where $F_{pos,\theta,\lambda}$ is the GWP feature at a specific position, orientation and frequency. $ stim $ is stimulus, $ G_{pos,\theta,\lambda,\phi} $ is the Gabor wavelet at a particular position, orientation, frequency and phase, and $ \cdot $ indicates dot product.

The dot product in the above equation makes it a linear filter model. But it's well established that neurons in the visual cortex behave like a nonlinear filter \cite{touryan_isolation_2002,sharpee_importance_2008}, for reasons such as saturation and so on \cite{sclar_coding_1990}. Here, we adopted a square root transformation to capture the nonlinearity:

\begin{equation}
	F = \sqrt{F_{pos,\theta,\lambda}}
\end{equation}
This nonlinear transformation was also applied to the other three encoding models.

After feature extraction, there are totally 10920 features for each image, which makes avoiding of overfitting the main concern in the feature mapping stage. Two measures were taken to address this problem. The first is to estimate a receptive field (RF) for each voxel, and only features located in the RF of a voxel were used for further activity prediction. The details of the RF estimation is the same as in \cite{kendrick_n._kay_identifying_2008}. The second is to fit the model with lasso regression which can yields very sparse weight values for large number of features \cite{tibshirani_regression_1996}.

\subsubsection{Gross local orientation model}
Too many features will lead to overfitting and worsen the prediction power of the encoding model on new samples \cite{yamashita_sparse_2008}. To reduce the number of features, we also proposed a simplified version of GWP model, which is called gross local orientation (GLO) model. The only difference between the GLO model and the GWP model is that the former takes the average of 8 orientation features at each location and spatial frequency as a gross local orientation feature, hence the number of features shrinks sharply to one eighth that of the GWP model.

\begin{equation}
F_{pos,\lambda} = \frac{1}{8}\sum_{\theta}\sqrt{\sum_{\phi}(stim \cdot G_{pos,\theta,\lambda,\phi})^2}
\end{equation}
where $F_{pos,\lambda}$ is the GLO feature at a specific position and frequency. $ stim $ is stimulus, $ G_{pos,\theta,\lambda,\phi} $ is the Gabor wavelet at a particular position, orientation, frequency and phase, and $ \cdot $ indicates dot product.

On the feature mapping stage, instead of estimating an RF to do feature reduction, we simply choose the feature that has the largest similarity (Pearson coefficient) to a voxel's responses as the only feature for predicting that voxel's activity.

\subsubsection{Retinotopy-only model}
To further simplify the GWP model, a retinotopy-only (RO) model was also adopted here. The RO model characterizes each voxel's activities as a function of the contrast and luminance of a specific region of visual images \cite{kendrick_n._kay_identifying_2008}. There are two input channels. One is the luminance channel which represents absolute deviation from mean luminance. The other is the contrast channel that represents the total energy contained in the image excluding overall luminance. Note that the RO model is invariant to the particular orientations and spatial frequencies present in the image.

To implement the RO model, first it need to choose metrics for luminance and contrast. Here we used a 2-D Gaussian envelope as the spatial weight function to calculate the local luminance and contrast for it's reasonable to presume that the receptive field of a voxel has spatial gradation such that portions of the image at the center of the receptive field contribute more strongly to the response than portions of the image at the periphery of the receptive field \cite{kendrick_n._kay_identifying_2008, dumoulin_population_2008}.

The spatially weighted luminance of a region is defined as:
\begin{equation}
L = \frac{\sum\limits_{i}w_i x_i}{\sum\limits_{i}w_i}
\end{equation}
where $L$ is the spatially weighted luminance, $w_i$ is the weight of pixel $i$, and $x_i$ is the luminance of pixel $i$. The spatially weighted contrast of a region is defined as:
\begin{equation}
C = \sqrt{\frac{\sum\limits_{i}w_i (x_i-L)^2}{\sum\limits_{i}w_i}}
\end{equation}
where $C$ is the spatially weighted contrast.

The RO model was applied to the same estimated receptive-field location as used for the GWP model. Both the RO model and GWP model were fit using the same lasso regression method.

\subsubsection{Zero model}
The zero model simply sets the feature to zero, thus excluding any information from the external stimulus. It's obvious that it is not an effective encoding model. 

\begin{equation}
F = 0
\end{equation}
where $F$ is the zero model feature.

The purpose of introducing the zero model is to verify whether the proposed framework with inner-state can still be used to encode and decode while the forward model contained in it has not any predictive ability.

% needed in second column of first page if using \IEEEpubid
%\IEEEpubidadjcol

\subsection{Natural image identification}
Identification is one kind of decoding, others forms of decoding include classification and reconstruction \cite{kendrick_n._kay_identifying_2008}. Identification takes two parameters as its inputs. One is a set of visual stimulus, the other is a measured brain activity pattern corresponding to one stimulus in the set. The work is to identify which stimulus in the image set corresponds to the given activity pattern.
%\begin{figure*}[htp]
%	\centering
%	\includegraphics[width=6.3 in]{decoding.eps}
%	\caption{Image identification using ISF and forward encoding model.}
%	\label{iden_frame}
%\end{figure*}

Identification could be accomplished by any kinds of encoding model. Fig. \ref{fig_framework_decoding} shows the procedure that uses ISF to identify the correct stimulus. First, for each image in the set, predict each voxel's activity using the forward model. All voxels' activities thus make up an activity pattern. Then the inner state model is applied to the predicted pattern and the measured pattern to yield an updated pattern which is the final prediction of ISF. Then a similarity value between each predicted pattern and the measured pattern is computed using Pearson correlation. The image whose predicted activity most closely matches the measured activity is chosen as the identification result.

The original validation set only contained 120 images. To investigate whether a decoder can handle much larger set of images, we measured identification performance for set sizes up to 1,000 images. The following procedure was used: first, a library of 1000 images was constructed. These images were randomly selected from the Internet and were different from the images used in the model estimation and image identification stages
of the experiment. Then, for set size $s$ and measured voxel activity pattern $\boldsymbol{m}_p$, identification performance was calculated as the probability that the predicted voxel activity pattern for the correct image is more correlated with $\boldsymbol{m}_p$ than the predicted voxel activity patterns for $s-1$ images drawn randomly from the library:
\begin{equation}\label{eq_pca}
f(\boldsymbol{m}_p,s) = \prod_{i=1}^{s-1}\frac{1000-g(\boldsymbol{m}_p)-i}{1000-i}
\end{equation}
\begin{equation}\label{eq_pca}
acc(s)=\frac{1}{120}\sum_{p=1}^{120}f(\boldsymbol{m}_p,s)
\end{equation}
where $f(\boldsymbol{m}_p,s)$ is identification performance under $\boldsymbol{m}_p$, $g(\boldsymbol{m}_p)$ is the number of library images whose predicted voxel activity patterns were more correlated with $\boldsymbol{m}_p$ than with the correct image and $acc(s)$ is the averaged identification performance over all measured voxel activity patterns under set size $s$.

\section{Experiments}

\subsection{Dataset}
We used the fMRI data set that was originally published in \cite{kendrick_n._kay_identifying_2008}. The stimuli set consisted of gray-scale natural images, content of which included humans, indoor and outdoor scenes, buildings, and food etc. All the images were down sampled, masked with 20-diameter circles and put on gray background. Images were presented in successive 4-s trials consisting of a 1 s presentation and 3 s gray background. Flashing technique was used in the 1 s presentation, that an image was flashed three times, like ON-OFF-ON-OFF-ON, where ON and OFF were both 200 ms respectively. Two subjects were used to collect data, with each of them participated five scan sessions. In each session, there were five model estimation runs, of which each run consisted of 70 different images presented two times, and two image identification runs, of which each run consisted of 12 different images presented 13 times. Totally, 1750 different images were used to train models and 120 different images were used to validate the performance of the trained models.

fMRI data were measured from occipital cortex at a temporal resolution of 1 Hz and a spatial resolution of 2 mm $\times$ 2 mm $\times$ 2.5 mm. The time series of BOLD were pre-processed to yield a response value for each voxel to each stimulus. Voxels were labeled with visual areas based on separately retinotopic mapping data using a multifocal mapping technique. There were 5512 (subject 1) and 5275 (subject 2) voxels in the early visual areas (V1, V2 and V3). The details of the experimental procedures and pre-processing methods are presented in \cite{kendrick_n._kay_identifying_2008}.

\subsection{Experimental results}
\subsubsection{Encoding}
\begin{figure}[htp]
	\centering
	\includegraphics[width=3.5 in]{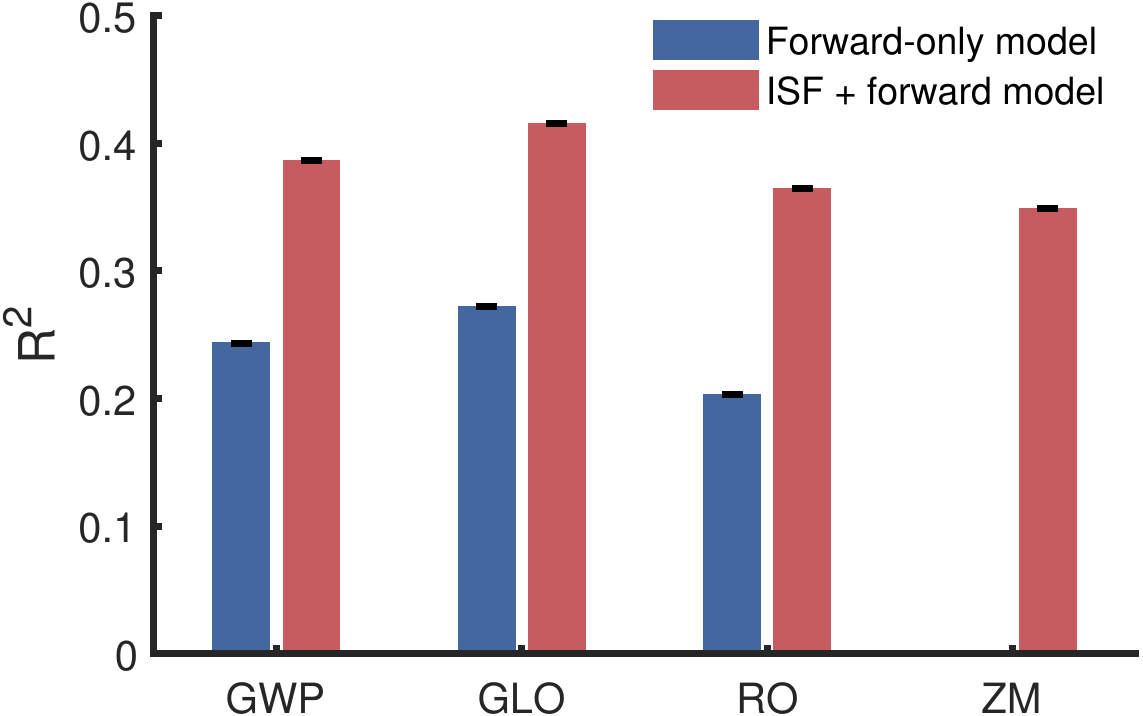}
	\caption{Comparison of encoding performance of the four forward encoding models and ISFs. The x-axis is the four forward models (blue bars) and ISFs combined with them (red bars), the y-axis is the mean $R^2$ across 2 subjects and the voxels that survived the $R^2$ threshold of 0.1. Error bars indicate $\pm1$ SEM across all the used voxels.}
	\label{fig_result_encoder_cmp}
\end{figure}

The encoding performance of the forward models and ISFs combined with them was defined as $R^2$ between the observed and predicted voxel responses to the 120 images in the validation set across the two subjects. For each forward model, as shown in Fig. \ref{fig_result_encoder_cmp}, the mean performance of ISF + forward model was found significantly higher than forward-only model (t test, $p < 0.01$ for all models). The GLO and ISF with GLO achieved the highest mean prediction $R^2$ of 0.28 and 0.42 in their respective model categories. The performance of RO ($R^2 = 0.20$) and ISF with RO ($R^2 = 0.36$) is lower than that of GWP ($R^2 = 0.24$) and ISF with GWP ($R^2 = 0.39$) respectively. The zero model has as expected no predictive power ($R^2 = 0$), but ISF with ZM still achieved a better predictive power of $R^2 = 0.35$.

\begin{figure*}[htp]
	\centering
	\includegraphics[width=6 in]{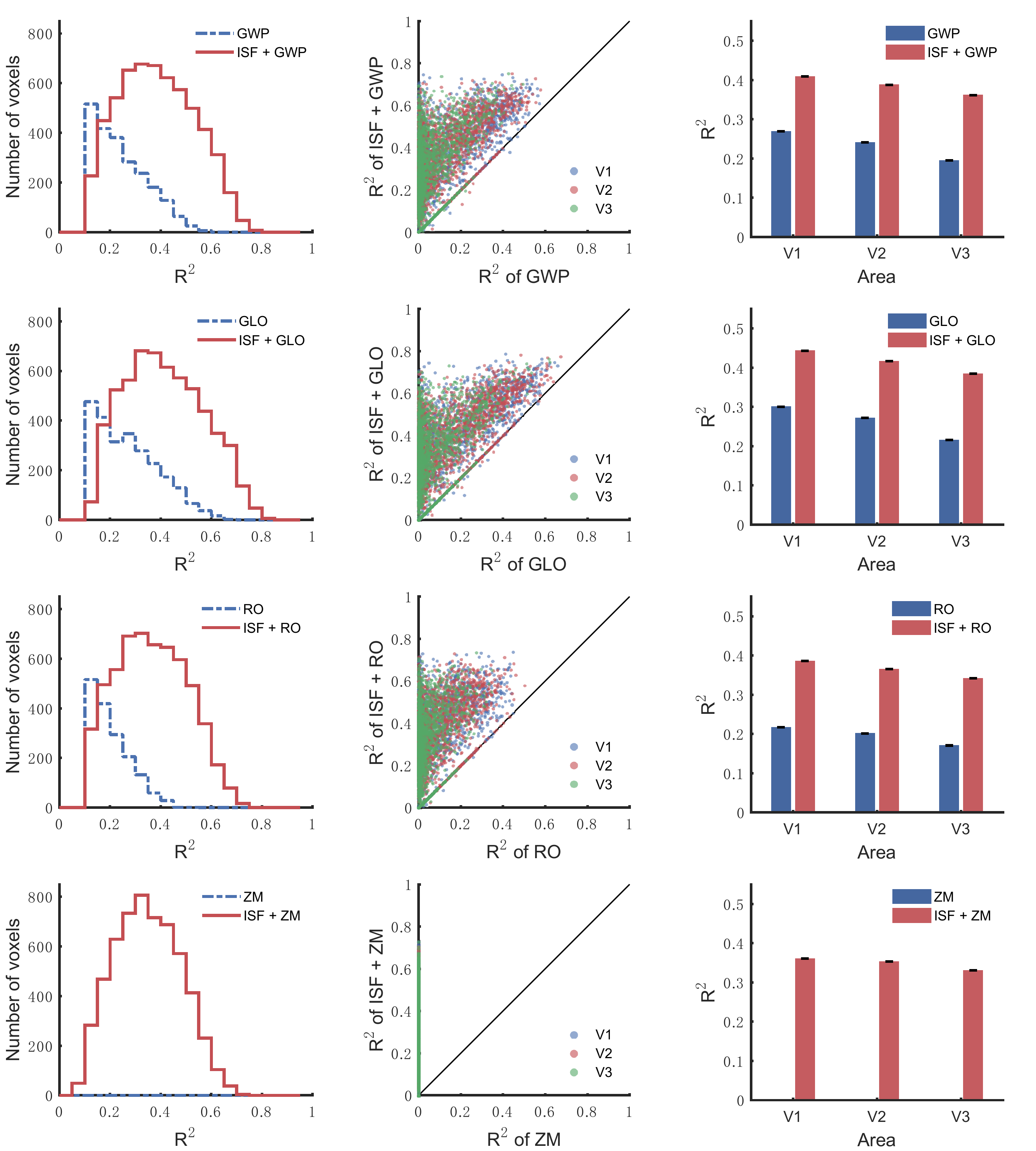}
	\caption{Encoding performance of the four forward models and ISFs combined with them. The encoding performance was defined as $R^2$ between the observed and
		predicted voxel responses to the 120 images in the validation set across the two subjects. Each row is the result of one model. The first column shows the prediction $R^2$ across the voxels that survived the $R^2$ threshold of 0.1; The second column shows the prediction $R^2$ in each
		voxel in the early visual cortex; the third column shows the mean prediction $R^2$
		across the voxels that survived the $R^2$ threshold of 0.1. Error bars indicate $\pm1$ SEM across all the voxels.}
	\label{fig_result_encoder}
\end{figure*}

Fig. \ref{fig_result_encoder} compares the performance of the four models across voxels in V1, V2 and V3. The amount of voxels that survived an $R^2$ threshold of 0.1 was larger for ISF with forward models than that of forward-only models (Fig. \ref{fig_result_encoder}, the first column). Improvements of the predictive power were found across voxels with their $R^2$ of forward-only model ranged from 0 to the maximum (Fig. \ref{fig_result_encoder}, the second column).  The mean $R^2$ of ISFs with forward models decreased from V1 to V3 (ISF + GWP: from 0.41 to 0.36; ISF + GLO: from 0.44 to 0.38; ISF + RO: from 0.39 to 0.34; ISF + ZM: from 0.36 to 0.33;). And the same trend was found on the mean $R^2$ of forward-only models (GWP: from 0.27 to 0.20; GLO: from 0.30 to 0.22; RO: from 0.22 to 0.17;) (Fig. \ref{fig_result_encoder}, the third column).

These results suggest that ISFs which combined the external stimuli with brain inner state can better explain voxels' fluctuations than forward-only models.

\begin{figure*}[htp]
	\centering
	\subfloat[]{
		\includegraphics[width=6 in]{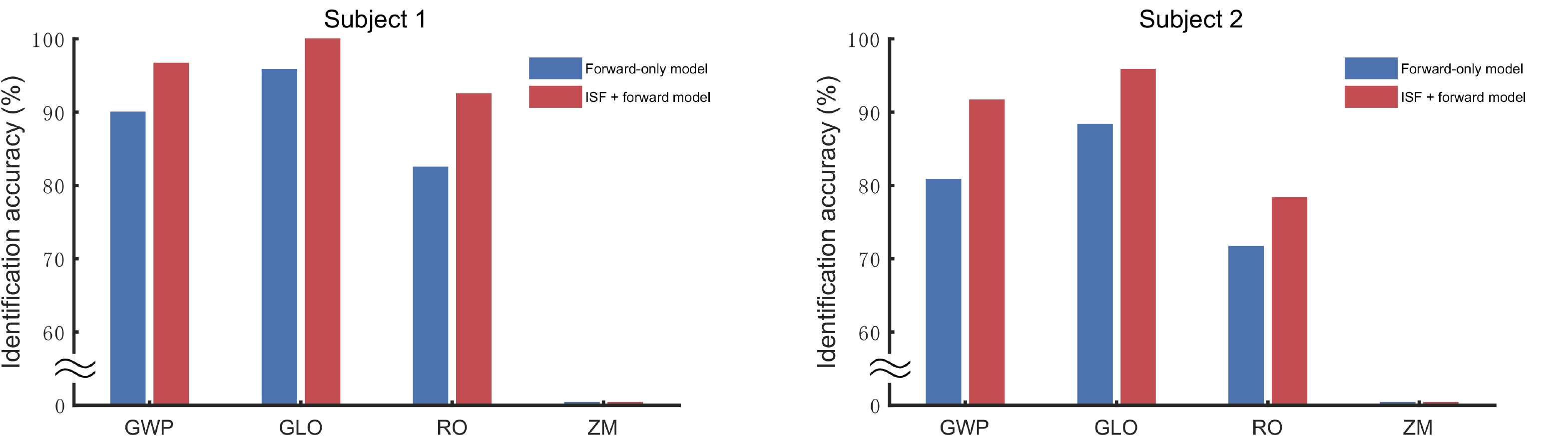}
		\label{fig_result_decoder_max}}
	\vfil
	\subfloat[]{
		\includegraphics[width=6 in]{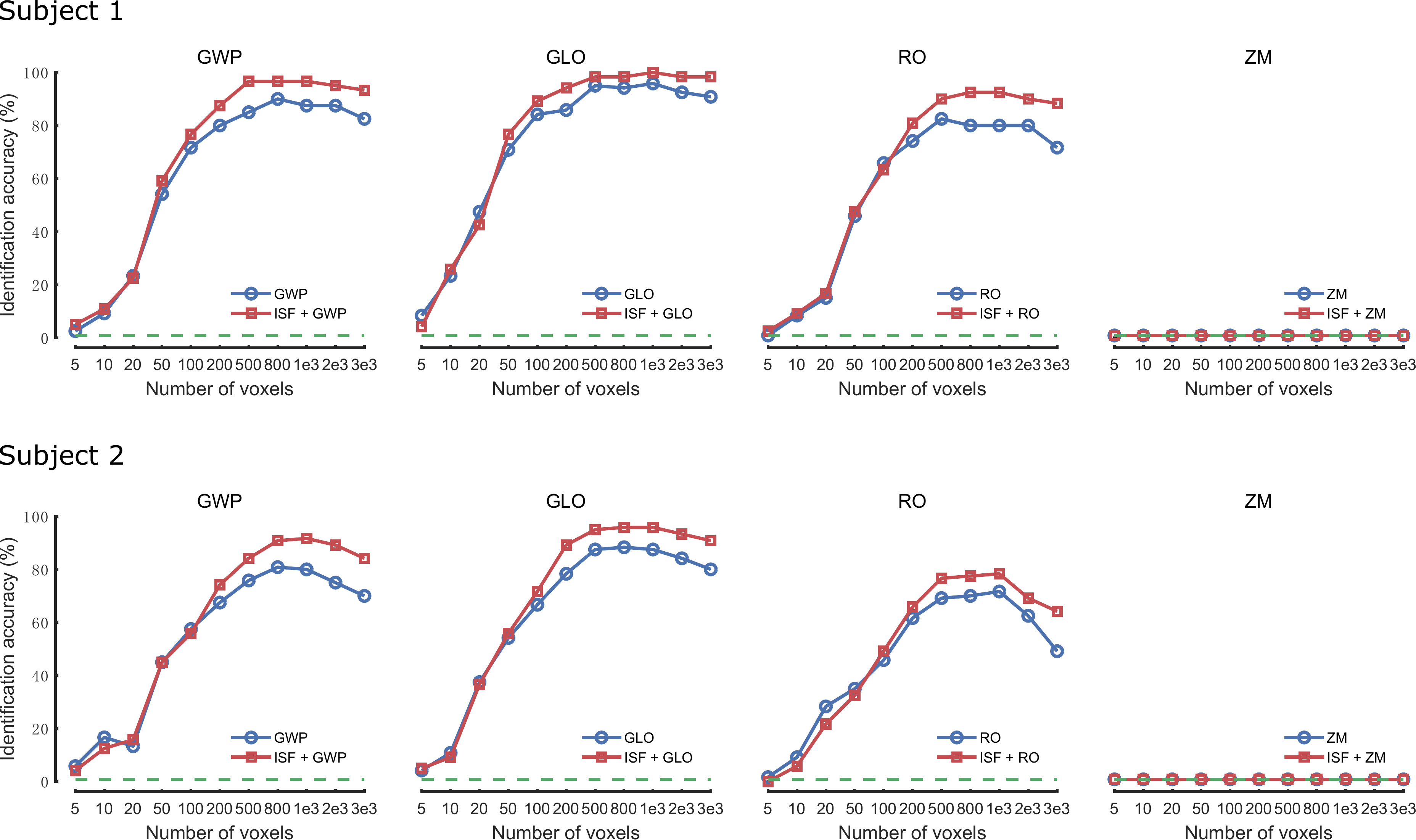}
		\label{fig_result_decoder_vox}}
	\caption{Decoding performance of the four forward models and ISFs combined with them. The decoding performance was defined as the accuracy of identifying the 120 images in the validation set. (a) Optimal performance of each model (blue bars) and ISF (red bars) for each subject. In all cases optimal performance was achieved using about 800 - 1000 voxels. (b) Effect of number of voxels on identification performance. Different numbers of voxels were selected according to their predictive power. The dashed green line indicates chance performance.}
	\label{fig_result_decoder}
\end{figure*}

\begin{figure*}[htp]
	\centering
	\subfloat[]{
		\includegraphics[width=6 in]{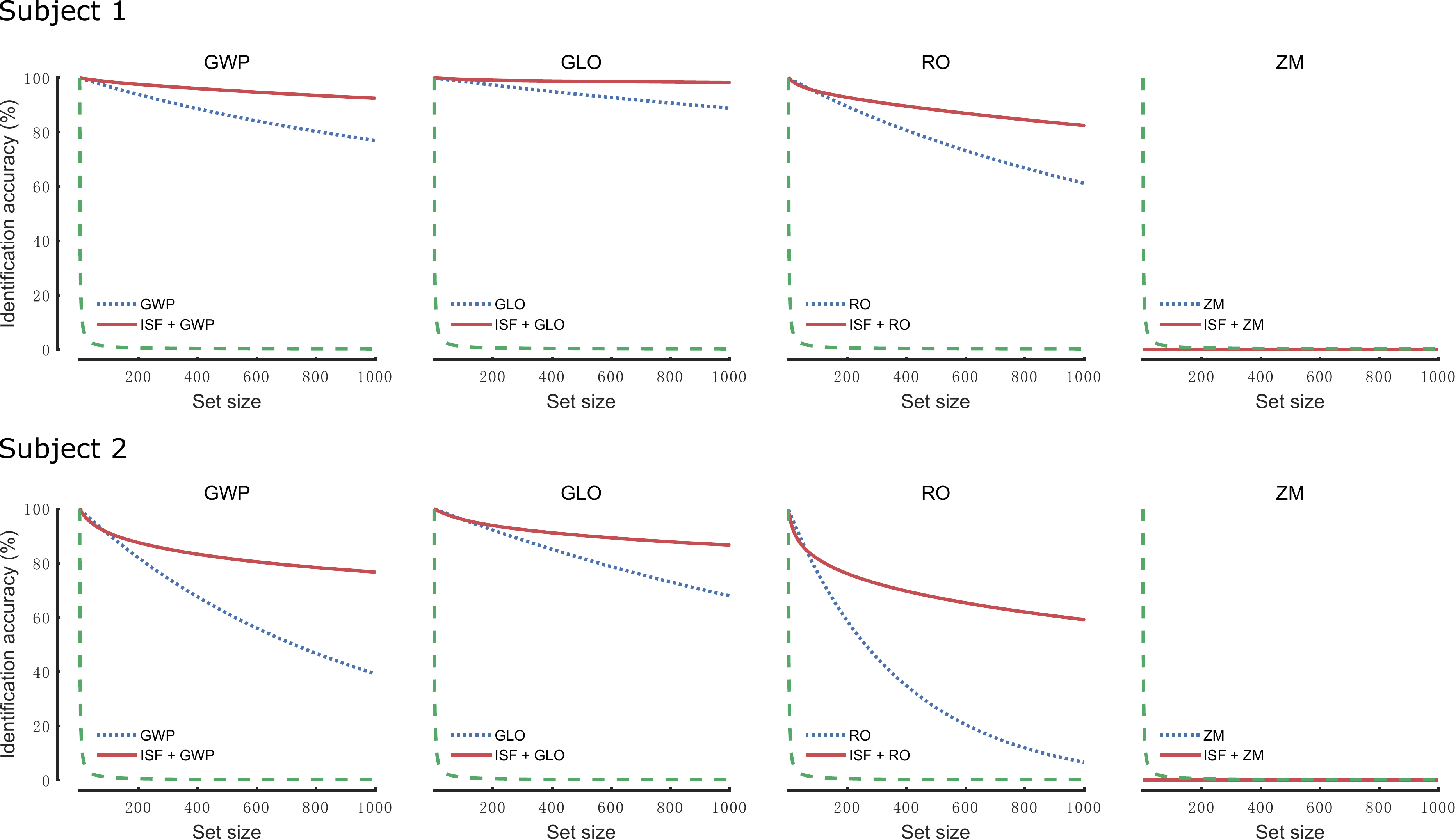}
		\label{fig_result_extdecoder_setsize}}
	\vfil
	\subfloat[]{
		\includegraphics[width=6 in]{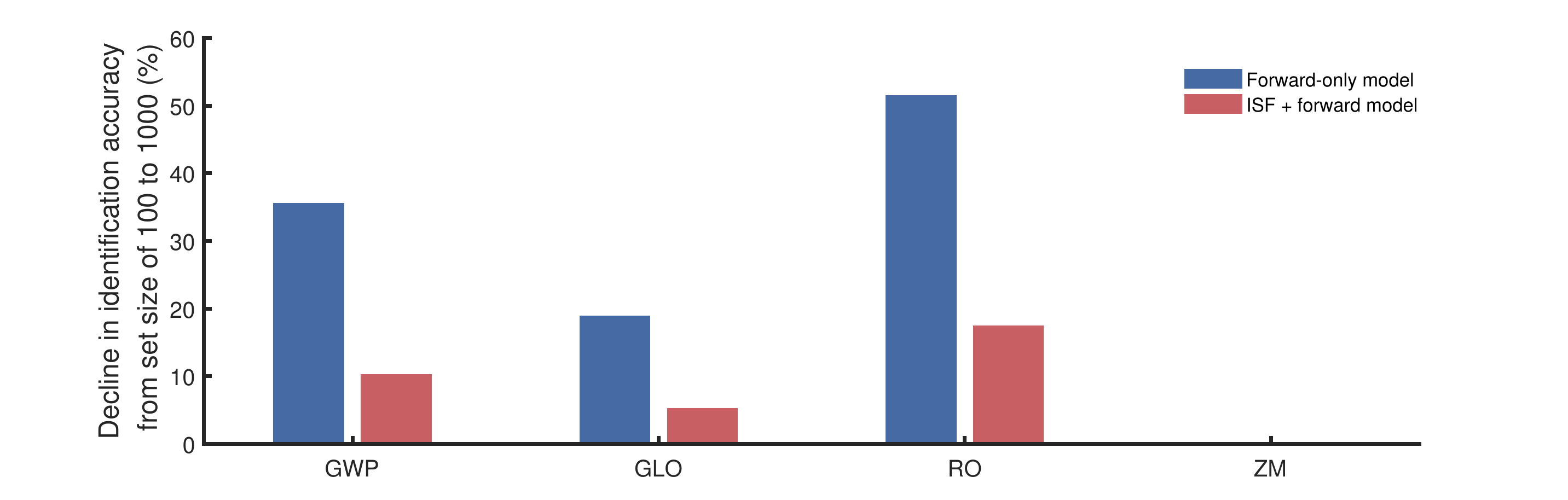}
		\label{fig_result_extdecoder_decline}}
	\caption{Scaling of identification performance with set size. (a) Performance curves of the four forward-only models and ISFs combined with them for the two subjects. The x-axis indicates size of the expanded image set, the y-axis indicates optimal identification performance using 800 voxels. The dashed green line indicates chance performance. (b) Comparison of the mean decline of different models in identification performance from image set of 100 to 1000 across the two subjects.}
	\label{fig_result_extdecoder}
\end{figure*}

\subsubsection{Decoding}
The decoding performance of the forward models and ISFs combined with them was defined as the accuracy of identifying the 120 images in the validation set. For subject 1, 90\%, 95.83\%, 82.5\% and 0.833\% of the images were identified correctly with the four forward-only models of GWP, GLO, RO and ZM respectively, while 96.67\%, 100\%, 92.5\% and 0.833\% of the images were identified correctly with ISFs with the four forward models respectively (Fig. \ref{fig_result_decoder_max}, left). For subject 2, identification accuracy was 80.83\%, 88.33\%, 71.67\% and 0.833\% with the four forward-only models respectively, while 91.67\%, 95.83\%, 78.33\% and 0.833\% with the four ISFs respectively (Fig. \ref{fig_result_decoder_max}, right). The chance level of identification accuracy is 0.833\%. Fig. \ref{fig_result_decoder_vox} shows how the identification performance varied with the number of voxels involved. For all the models (except for the ZM, whose performance stayed at the chance level of 0.833\%), the performance first increased and then declined as the number of voxels increased. In all cases optimal performance was achieved using about 800 - 1000 voxels.

To investigate whether these models can be able to handle much larger sets of images, we also measured identification performance on an extended image set with set sizes up to 1000 images (Fig. \ref{fig_result_extdecoder_setsize}). As set size increased from 100 to 1000, for subject 1, identification performance of the three forward-only models (GWP, GLO and RO) declined from 96.83\%, 98.67\% and 94.48\% to 77\%, 88.9\% and 61.2\% respectively, while that of ISFs with the three forward models declined from 98.61\%, 99.48\% and 95.04\% to 92.5\%, 98.33\% and 82.5\% respectively. For subject 2, identification performance of the three forward-only models (GWP, GLO and RO) declined from 90.41\%, 95.96\% and 76.33\% to 39.2\%, 67.9\% and 6.6\% respectively, while that of ISFs with the three forward models declined from 91.05\%, 95.99\% and 81.51\% to 76.67\%, 86.67\% and 59.17\% respectively. The performance of the ZM stayed at 0 for all set sizes. As shown in Fig. \ref{fig_result_extdecoder_decline}, the mean performance decline of ISFs with forward models (for GWP, GLO and RO was 10.25\%, 5.23\% and 17.44\% respectively) was significantly lower than that of the forward-only models (for GWP, GLO and RO was 35.52\%, 18.92\% and 51.51\% respectively).

These results suggest that ISF which combined the external stimuli with brain inner state can be more effectively exploited in brain decoding than forward-only model. In addition, although ISF without effective information from the external world can still achieve better encoding performance, it doesn't have the ability to do image identification.

\section{Discussion}
The present work addresses the problem that traditional fMRI encoding models ignore abundant information from brain inner state. We proposed a novel encoding framework (ISF) that combined information from the external world with information from brain inner state. The ISF includes two parts: a forward encoding model that captures responses to stimuli and an inner state model that captures activities evoked by other relative voxels. The encoding model included in the ISF is replaceable, which makes the ISF very flexible. Using a set of real experimental data and four different forward encoding models, we demonstrated that ISFs could better explain voxels' fluctuations than forward-only models, and ISFs combined with effective forward models achieved state-of-the-art identification performance with the best accuracy being 100\%.

\subsection{Explanation of the brain inner state model}
The brain is a complex state machine whose sensory cortex is constantly receiving information from the external world as well as other parts in the brain. At present, all encoding models only focus on improving the feature extraction method, and simply discard the prediction residuals for they are considered noise, such as physical noise and physiological noise. We found that voxel responses explained by these forward models remained low level no matter how complex these models were, and there were correlations among the predicted residuals of voxels which were difficult to explain only by noises. Based on these facts, we hypothesized that the prediction residuals of the forward model reflects the brain inner state, and built a linear model for the inner state. This data-driven model of the brain inner state has several features. First, the estimated inner state is linearly independent to image features. This is guaranteed by the orthogonality between residuals and independent variables in a linear model. Therefore, there is no local information of images in the estimated inner state. Second, the inner state is estimated using connectivities between voxels, which has its own neural basis, such as default mode network \cite{greicius_functional_2003, greicius_resting-state_2009}, lateral connections in the visual cortex \cite{smith_nonstimulated_2010, douglas_neuronal_2004}, and top-down modulated feedback connections from high level regions \cite{gilbert_top-down_2013}. Hence, the inner state may reflect the intrinsic connections in the brain, such as structural and functional connectivity \cite{van2010exploring, thomas2011organization}, and contain global information from other brain areas \cite{angelucci_circuits_2002}, such as attention and expectation \cite{summerfield_expectation_2009, kok_less_2012}.

\subsection{Identification performance with different pattern similarity criteria}
The identification is carried on by comparing activity patterns of predicted with that of measured. The measurement of pattern similarity adopted in this work and in other fMRI encoding researches \cite{kendrick_n._kay_identifying_2008,naselaris_voxel-wise_2015,guclu_unsupervised_2014,gucclu2015deep} was Pearson correlation coefficient which can be viewed as the cosine of the angle between two patterns. However, the Euclidean distance is also an important measurement for pattern similarity. Euclidean distance is more restrictive since the most similar pattern must be identical to the given pattern with it as the criterion. Here, we compared the identification performance with both Pearson's $r$ and Euclidean distance as the measurement of pattern similarity under forward-only model and ISF.

\begin{figure}[htp]
	\centering
	\includegraphics[width=3.5 in]{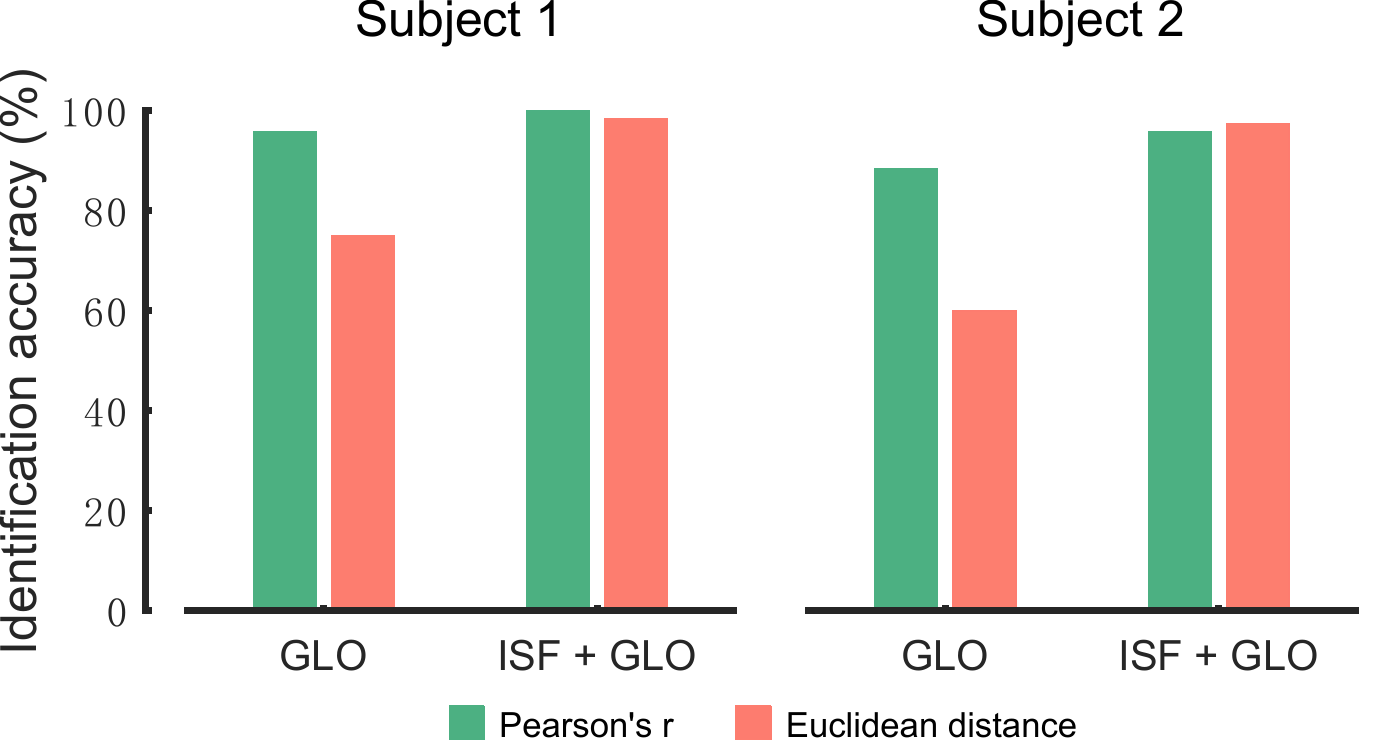}
	\caption{Comparison of identification performance of GLO and ISF with pattern similarity measurement of Pearson's $r$ and Euclidean distance.}
	\label{fig_discussion_decoder_mse}
\end{figure}

As shown in Fig. \ref{fig_discussion_decoder_mse}, for subject 1, as the measurement of pattern similarity changed from Pearson's $r$ to Euclidean distance, the identification accuracy of ISF with GLO was almost constant (from 100\% to 98.33\%), while that of GLO dropped a lot (from 95.87\% to 75\%). The same trend was also found in subject 2. Comparison using other forward models and ISFs yielded the same result (data were not shown here). This demonstrated that the predicted activity pattern using ISF is closer to the measured activity pattern than that using forward-only model, hence the decoding performance is more robust with different pattern similarity criteria.

\subsection{Why ISF achieved better identification performance?}
Considering that the inner state model alone, as in the Zero model, can not improve the identification accuracy, then why it works when combined with any effective forward model is a question that needs an intuitive explanation. The inner state model captures some intrinsic structure in the prediction residuals of forward model, which is independent to the external stimuli. For a given measured activity pattern which is comprised of three components: component that can be predicted by the stimulus, component that can be predicted by the inner state, and noise, the residual of forward model of the correct image is more consistent with the inner state than that of the incorrect image, even when the mean squared residual (MSR) of the incorrect image is smaller. Identification with a forward-only model selects the image that has the smallest MSR as the result, while identification with ISF considers not only the amount of MSR, but also how much of the residual of the forward model can be explained by the inner state. In the case that the MSR of the correct image is not the smallest, forward-only model will inevitably make an incorrect choice, howerver, when combined with the inner state model, ISF still has a chance to make a better choice.

\subsection{Future works}
The inner state of a voxel is estimated by the intrinsic connectivity which is independent to responses evoked by extern stimuli. In this work, the estimation was completed by applying PCA to the residual matrix. In addition to PCA, independent component analysis (ICA) and cross-correlation analysis (CCA) are also general tools for detecting functional connectivity under resting-state \cite{mckeown_independent_2003,ma2007detecting}, hence can be used in our ISF. One of the future works should be to compare these component analysis approaches to discover a more appropriate tool for the estimation of intrinsic connectivity under task-state.

Although we focused on fMRI encoding and decoding, the idea of ISF can be extended to apply to the pattern identification of other neuroimaging signals such as EEG and MEG. In particular, its application to the brain computer interface (BCI) would be of great interest.

% if have a single appendix:
%\appendix[Proof of the Zonklar Equations]
% or
%\appendix  % for no appendix heading
% do not use \section anymore after \appendix, only \section*
% is possibly needed

% use appendices with more than one appendix
% then use \section to start each appendix
% you must declare a \section before using any
% \subsection or using \label (\appendices by itself
% starts a section numbered zero.)
%

%\appendices
%\section{Proof of the First Zonklar Equation}
%Appendix one text goes here.

% you can choose not to have a title for an appendix
% if you want by leaving the argument blank
%\section{}
%Appendix two text goes here.

% use section* for acknowledgment
\section*{Acknowledgment}
This work was supported by the 973 Program (No. 2015CB351703).

% Can use something like this to put references on a page
% by themselves when using endfloat and the captionsoff option.
\ifCLASSOPTIONcaptionsoff
  \newpage
\fi

% trigger a \newpage just before the given reference
% number - used to balance the columns on the last page
% adjust value as needed - may need to be readjusted if
% the document is modified later
%\IEEEtriggeratref{8}
% The "triggered" command can be changed if desired:
%\IEEEtriggercmd{\enlargethispage{-5in}}

% references section

% can use a bibliography generated by BibTeX as a .bbl file
% BibTeX documentation can be easily obtained at:
% http://mirror.ctan.org/biblio/bibtex/contrib/doc/
% The IEEEtran BibTeX style support page is at:
% http://www.michaelshell.org/tex/ieeetran/bibtex/
\bibliographystyle{IEEEtran}
% argument is your BibTeX string definitions and bibliography database(s)
\bibliography{mybib.bib}
\end{document}